# Design and Implementation of Scalable Communication Interfaces for Reliable and Stable Real-time Co-Simulation of Power Systems

Qi Xiao, Jongha Woo, Lidong Song, Ning Lu
NC State University, Raleigh, NC, USA
{**qxiao3**, nlu2}@ncsu.edu

Victor Paduani
New York Power Authority, Albany, NY, USA

*Abstract*—Co-simulation offers an integrated approach for modeling the large-scale integration of inverter-based resources (IBRs) into transmission and distribution grids. This paper presents a scalable communication interface design and implementation to enable reliable and stable real-time co-simulation of power systems with high IBR penetration. The communication interface is categorized into two types: local and remote. In local scenarios, where subsystems are connected within a single *local area network* (LAN), low-latency communication facilitates the seamless integration of electromagnetic transient (EMT) and phasor-domain models, enabling efficient interactions with power and energy management algorithms. For remote scenarios, data exchange is achieved via internet-based file sharing or VPN-enabled communication. The performance of both methods is evaluated using OPAL-RT as a real-time simulator, demonstrating scalability, effectiveness, and challenges specific to real-time co-simulation applications. To mitigate instability arising from data resolution mismatches in time-sensitive co-simulations, a real-time data extrapolation method is proposed. This approach significantly enhances stability and reliability, ensuring more accurate simulation outcomes. The implementation code is available on GitHub, providing researchers the tools to replicate and expand upon this work.

## I. INTRODUCTION

The large-scale integration of distributed energy resources (DERs) and inverter-based resources (IBRs) has significantly increased the scale and complexity of power systems [1]. This has necessitated electromagnetic transient (EMT) simulations at microsecond resolution to ensure grid stability, as well as integrated transmission and distribution (T&D) simulations capable of addressing both system levels. However, such simulations require substantial computational resources, often surpassing the capabilities of a single modeling platform designed for either EMT or phasor-domain simulations. Furthermore, IBR and DER power and energy management algorithms frequently rely on external solvers operating independently from the power system simulation platforms.

In recent years, co-simulation has become a powerful approach for enabling the concurrent operation of multiple subsystems within a unified environment. It supports the integration of diverse modeling environments and control frameworks, enabling the exchange of monitoring signals and control commands to simulate complex subsystem interactions accurately [2]. However, while co-simulation offers advantages such as flexibility and scalability, it also presents challenges, particularly in managing latency and synchronization when coordinating subsystems with varying time steps, sampling rates, and control intervals. Overcoming these challenges is crucial for ensuring simulation accuracy and maintaining system stability.

In our previous studies, we observed that the choice of co-simulation communication interface is highly dependent on the application. For instance, in [3], we investigated device-level co-simulation to integrate models like EMT and phasor models. Meanwhile, in [4]-[6], we focused on network-level co-simulation frameworks to analyze interactions resulting from DER integration. Co-simulation with energy management systems is also crucial for replicating real-world conditions [7], [8], though it inherently involves communication latencies. Effectively addressing latencies with tailored data exchange methods is crucial, especially for applications like cybersecurity analysis.

To address these challenges, this paper presents solutions developed for various co-simulation scenarios and introduces a suite of communication interfaces that enable sharing real-time simulator (RTS)-based testbeds developed at North Carolina State University (NCSU) with other institutions for collaborative real-time simulations. The communication interfaces are categorized into two types: *local* and *remote*.

Local co-simulation connects subsystems within a shared environment, typically over a local area network (LAN), enabling low-latency communication. This setup is ideal for device-level simulations within transmission or distribution networks or for integrating power networks with local control systems. In contrast, remote co-simulation facilitates collaboration among geographically dispersed teams and systems across institutions (e.g., university campuses) via internet connections, addressing challenges posed by distance and network constraints. To support both time-sensitive and time-insensitive remote co-simulations, we develop file-sharing mechanisms and VPN-based TCP/IP socket communication. Our findings indicate that data resolution mismatches in time-sensitive co-simulations can cause instability in IBRs. To mitigate this issue, we propose a real-time data extrapolation method, significantly enhancing stability and reliability, leading to more accurate simulation results.

The performance is evaluated using OPAL-RT as the RTS, demonstrating the scalability and challenges of real-time co-simulation, as well as the effectiveness of the proposed extrapolation methods. To facilitate researchers to replicate and expand upon our work, we published our code for the communication interface designed for both *local* and *remote* on our GitHub repository [9].

This research is supported by the U.S. Department of Energy's Office of Energy Efficiency and Renewable Energy (EERE) under the Solar Energy Technologies Office, Award Number DE-EE0008770 and NSF ECCS, PD 18-7607 Energy, Power, Control, and Networks (EPCN), Award Number (FAIN): 2329536.

## II. METHODS

This section details the communication interface setup for the local and remote real-time co-simulation testbed, using a microgrid simulation testbed as an illustrative example.

### A. Local Communication Interface Design

We define *local co-simulation* as configurations where all subsystem models are connected within a single LAN, as illustrated within the dotted-line box in Fig. 1. In this setup, a microgrid control system (MCS) modeled on a networked personal computer (PC) can interact with a microgrid testbed modeled on a RTS using the simulated Modbus communication protocol over a LAN.

The microgrid testbed on the RTS platform can be further categorized into two model types: the EMT domain and the phasor domain. For example, as shown in Fig. 1, high-power inverter-based resources (IBRs), including a 3 MW PV farm, a 2 MWh battery energy storage system (BESS), a 1 MVA diesel generator (DG), and their device-level controllers, are modeled in the EMT domain using eMEGASIM with a 100-µs timestep. This approach captures microgrid frequency and voltage transients, enabling the evaluation of different IBR control modes, such as grid-forming and grid-following. Simultaneously, the IEEE 123-bus distribution network and associated loads are modeled in the phasor domain using ePHASORSIM with a 1-ms timestep. To ensure synchronized data exchange between the EMT and phasor domains, phasor-domain signals are used to update EMT waveforms through a first-order low-pass filter (LPF) with delay compensation, eliminating the need for time interpolation. For more details on this approach, please refer to [3].

The MCS manages energy and power balancing for the microgrid at intervals of up to 5 minutes. Within each 5-minute interval, balancing actions are handled by device-level controllers for BESS, DG, and MW-level PV, ensuring instantaneous microgrid frequency and voltage control. Thus, these device-level controllers are simulated on RTS using eMEGASIM alongside the IBRs, ensuring that control actions are executed without delay.

However, as the MCS receives measurement data (see the purple-line box in Fig. 1) and sends control commands (see the red-line box in Fig. 1) to the microgrid testbed via LAN, data exchange delays may occur. These delays, influenced by factors such as MCS computing time, data exchange rates, and communication methods, must be effectively managed to avoid misaligned measurements, which could lead to erroneous decisions by the MCS.

The local communication interface, hosted on another PC, enables real-time communication between the RTS and the MCS running on the PC. Using the Modbus protocol over Ethernet, the interface retrieves measurements from RTS and forwards it to the MCS via TCP/IP sockets. Concurrently, it can send control commands from the MCS to the RTS testbed for execution. All measurements and control commands are time-stamped and logged in the interface's database, ensuring accurate synchronization and traceability.

Figure 2 illustrates the workflow of the communication interface design. In this setup, the RTS operates as the Modbus server, while the MCS acts as the TCP/IP server. The local communication interface serves as the client for both protocols. During initialization, the TCP/IP address and port number are configured to establish communication links between the RTS, the MCS, and the local interface.

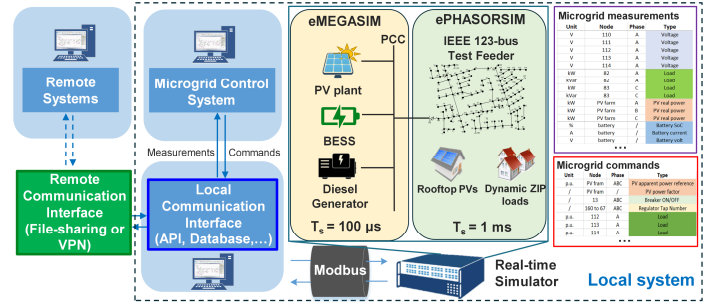

Fig. 1. Layout of the local and remote real-time co-simulation testbed.

Once the simulation begins, the interface performs the following tasks: 1) collect time-stamped measurements from RTS via Modbus at a defined polling frequency; 2) log the data in the database and synchronize the system clock using the embedded simulation time, which acts as the global clock for the co-simulation testbed; 3) transmit measurements to the MCS at a predefined interval for generating control commands; and 4) receive time-stamped control commands from the MCS, log them in the database, and forward them to the RTS for execution. To ensure seamless communication and prevent command loss, the interface utilizes Python-based multi-threading, enabling concurrent handling of upstream measurement collection (RTS to MCS) and downstream command transmission (MCS to RTS).

Figure 3 depicts the time sequence of the local co-simulation testbed. In the EMT and phasor co-simulation model on the RTS, the system propagation delay ($t_{d1}$) can reach up to $2T_{s,phasor} + T_{s,EMT}$, which corresponds to 2.1 ms in this configuration [2]. Additionally, the delay ($t_{d2}$) between RTS and MCS includes: 1) round-trip communication time ($\Delta t_{d4} + \Delta t_{d6} + \Delta t_{d8} + \Delta t_{d10}$); 2) data processing time in the local communication interface ($\Delta t_{d5} + \Delta t_{d9}$); and 3) control action calculation time at the MCS ($\Delta t_{d7}$). Since the MCS operates on a timescale of up to 5 minutes, the millisecond-level communication latency is negligible and does not affect system performance when properly managed.

With this communication setup, synchronized co-simulation is achieved by effectively interfacing the microgrid system-level controller with the device-level controllers residing on field devices. This method is readily applicable to co-simulation scenarios involving multiple power and energy management systems interacting with a single or interconnected network of real-time simulation testbeds within the same LAN.

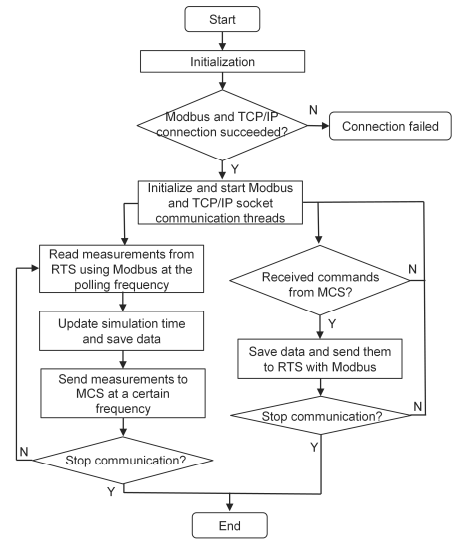

Fig. 2. Workflow of the local communication interface in the co-simulation testbed.

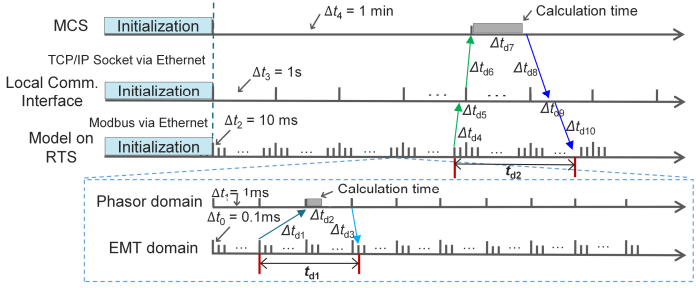

Fig. 3. Time sequence of the local co-simulation testbed.

### B. Remote Communication Interface Design

Remote co-simulation distributes subsystems on separate simulation platforms connected via internet rather than a LAN. This setup requires an improved and secure communication interface design to address challenges such as high latency and limited bandwidth. Applications requiring real-time control actions (e.g., tripping protection relays) demand low latency, whereas non-critical actions (e.g., dispatching generator setpoints) can tolerate higher delays. To accommodate these differing latency requirements, we propose two approaches: *file-sharing* (see Fig. 4(a)) and *VPN-based communication* (see Fig. 4(b)).

#### 1) File-Sharing Approach

The file-sharing method uses platforms such as cloud-based storage services to exchange data between subsystems. Each subsystem writes its local data to a shared file and retrieves remote data for simulation. As shown in Fig. 4(a), the workflow involves two steps: 1) write local data to the shared file, synchronized via the cloud service and 2) retrieve synchronized data from other subsystems, incorporating it into simulations, and updating the shared file with new data. This process is automated across all communication interfaces. To maintain synchronization and data alignment, all exchanged data are time-stamped.

Latency in this method can be influenced by many factors (e.g., local processing time, file size, internet speed, and cloud storage synchronization delays). For files less than a few megabytes, synchronization typically ranges from 5 to 30 seconds but can increase sharply with larger files or slower connections. Figure 5(a) depicts the time sequence of a communication cycle, from initial data transmission ($t_{g0}$) to the receipt of all updates ($t_{g12}$), with cycle duration scaling with the number of participating subsystems. Similar to the local communication interface, mechanisms such as scheduling, locking, or utilizing multiple shared files can be applied to address issues like data overwrites or conflicts during simultaneous access. However, it is important to note that these measures may introduce additional latency.

#### 2) VPN-based Approach

The VPN-based approach creates a secure, encrypted virtual private network to emulate a local network over the internet. This enables synchronized, low-latency data exchange, making it ideal for time-sensitive simulations while offering stronger security than WebSocket, MQTT, or public IP setups with port forwarding. Each subsystem's communication interface employs multi-threaded processes and non-blocking sockets for efficient concurrent data handling. As shown in Fig. 4(b), direct communication replaces file synchronization, significantly reducing latency compared to the file-sharing approach.

Figure 5(b) shows the time sequence of VPN-based communication, which enables faster data transfer for real-time control signals. However, it remains slower than local co-simulation due to inherent encryption, decryption, and internet propagation delays. Latency increases with the number of remote systems, potentially requiring additional resources to manage multiple connections.

While VPN ensures low latency, high security through strict access control, and a reliable connection, its implementation often requires lengthy administrative approvals and firewall reconfigurations. This makes VPN well-suited for real-time control applications involving sensitive utility models and data. In contrast, the file-sharing approach, though less secure and slower, is simpler to set up and supports multiple external users, making it ideal for testing power and energy management applications without sensitive data or models.

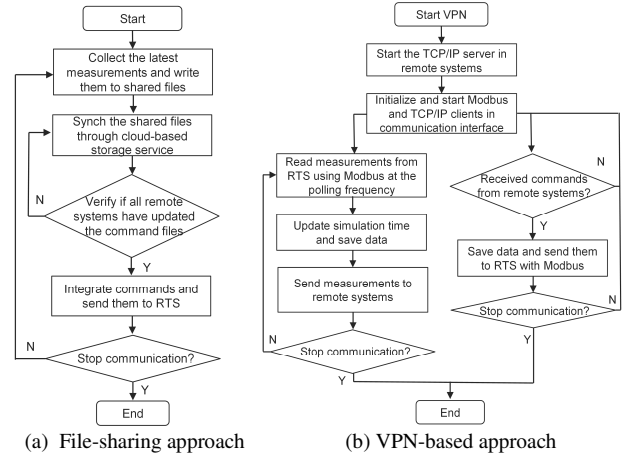

(a) File-sharing approach     (b) VPN-based approach

Fig. 4. Workflow of the remote communication interface.

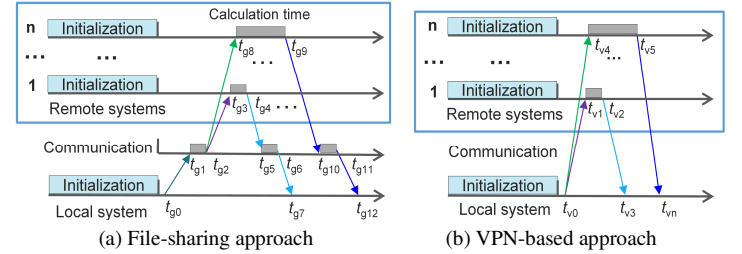

(a) File-sharing approach     (b) VPN-based approach

Fig. 5. Time sequence of the remote communication interface.

### III. SIMULATION RESULTS

This section presents simulation results to demonstrate the local and remote communication interface design, using OPAL-RT as the RTS.

#### A. Local Communication Interface Testing

To evaluate the accuracy of measurements recorded by the communication interface, data were collected from the microgrid testbed during the sequential energization of five distribution load groups (LGs) at 100-second intervals. Measurements were sampled from the OPAL-RT testbed at 1-second intervals. Figure 6 presents a comparison between the simulated PCC voltage, frequency, and system power recorded by OPAL-RT at a high resolution of 1 ms with the corresponding 1-second sampled data obtained through the local communication interface.

The results show a close alignment between the power response recorded by the communication interface and the real-time OPAL-RT data. This consistency is due to the relatively slow dynamics of power responses compared to the faster voltage and frequency variations. These findings confirm that the communication interface maintains sufficient fidelity for real-time control and monitoring applications.

To assess the stability of the communication interface, an SOC-based control strategy was implemented on the MCS, with the simulation running for twenty consecutive days. Results, as discussed in [10], demonstrate that the interface remains stable over extended durations. Due to space limitations, the detailed results are not included here.

Since the MCS operates on a minute-level control interval, local network communication and data processing—typically in the millisecond range—are negligible and do not impact control performance.

This case study demonstrates that the local communication interface remains stable and supports real-time data exchange and control in long-duration simulations.

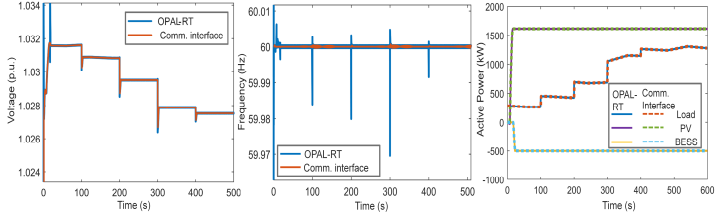

Fig. 6. Comparison of PCC voltage, frequency, and power measurements recorded by OPAL-RT and the local communication interface.

### B. Remote Communication Interface Testing

For the remote real-time co-simulation, we tested a setup involving two subsystems located at NCSU and another institute. Each subsystem comprised an OPAL-RT simulator, a local communication interface, and an MCS.

In the *file-sharing* approach, we implemented a load-following control use case to evaluate the interaction between the two remote systems. Both OPAL-RT simulators ran an IEEE 123-bus system, with the remote system's nodes following the load power of the corresponding nodes in the NCSU system. No control commands were sent back from the remote system to NCSU. At NCSU, approximately 500 measurements (4 KB each) were updated every 60 seconds via Google Drive. The remote system continuously monitored file timestamp changes and used the updated data as load power references.

Figure 7 shows the total load profile for both systems and the observed delay distribution. Synchronization speed varied due to multiple factors, resulting in delays of 1 to 8.5 seconds between transmitting and receiving node power references, with most delays falling within 1.5 to 4 seconds. Load power references at the remote system were updated approximately every 60 seconds due to file synchronization timing. If bidirectional communication were required (e.g., sending control commands from the remote system to NCSU), this delay would nearly double. Consequently, the file-sharing approach is best suited for applications where data exchange intervals far exceed the inherent latency, ensuring data integrity, system stability, and minimizing the risk of data loss.

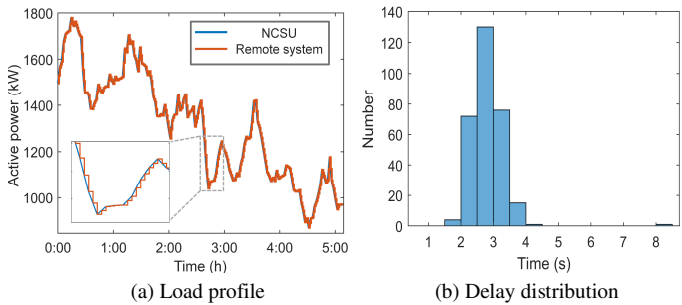

(a) Load profile    (b) Delay distribution

Fig. 7. Simulation results of file-sharing approach.

In the *VPN-based* configuration, the remote system connects to NCSU's local network via VPN, with the remote system functioning as the TCP/IP server and NCSU as the client. Testing revealed a communication latency of approximately 20 ms between the two local communication interfaces.

This setup enables a transmission and distribution (T&D) co-simulation, where the phasor-domain IEEE 118-bus transmission system model (with a 10-ms resolution) operates on the RTS at the remote system, while the IEEE 123-bus distribution system model with EMT-domain IBRs runs on NCSU's RTS [11]. The distribution system connects to Bus 102 of the transmission system via its point of common coupling (PCC). During co-simulation, the transmission system provides voltage magnitude, phase angle, frequency, and feeder switch signals to the distribution system, forming an equivalent voltage source. Conversely, the distribution system sends feeder header active and reactive power data back to the transmission system. Ensuring the accuracy and precision of the exchanged data is critical, as any processing or communication errors can degrade co-simulation performance. To guarantee data alignment, all exchanged data are time-stamped.

To evaluate the dynamic performance of the T&D co-simulation, a five-cycle three-phase-to-ground fault was applied at Bus 94 of the transmission system. Figure 8(a) presents the voltage response in both systems. Results indicate that due to computational and communication delays, the PCC voltage data received by the distribution system lags approximately 37–55 ms behind the actual voltage observed in the transmission system. Additionally, while the transmission system updates data at a 10-ms resolution, the data received at NCSU is updated every 17–35 ms, influenced by local communication and computation delays.

The received voltage and frequency data are transformed into a PCC voltage source within the EMT model, which is connected to PV and BESS systems. These IBRs synchronize with the PCC voltage using a phase-locked loop (PLL). Figure 8(b) compares the frequency data received from the transmission system with the frequency measured by PLL. The PLL-measured frequency shows significant spikes and oscillations due to a resolution mismatch, as the received frequency data has a coarser resolution than the EMT model's 100 μs precision.

This mismatch can disrupt synchronization in grid-connected IBRs, potentially leading to instability, particularly when co-simulation subsystems exhibit significant resolution disparities. To ensure system stability and improve the reliability of T&D co-simulations, it is essential to align data exchange resolutions or implement data smoothing techniques.

While methods such as LPF are commonly employed to smooth received data in local distribution co-simulations, LPF can introduce substantial delays and errors when there are large resolution differences. To mitigate this issue, we propose a real-time data extrapolation method, specifically designed for VPN-based co-simulation scenarios with significant resolution disparities, as described in Equations (1)–(3).

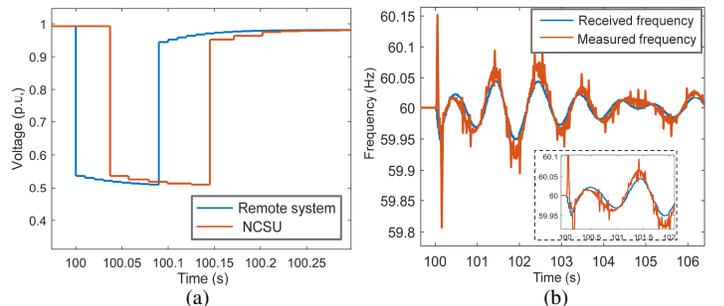

(a)    (b)

Fig. 8. (a) Comparison of the sent and received PCC voltage between the remote system and NCSU. (b) Comparison of the received and PLL-measured frequency at NCSU.

$$y_t^e = y_{t-1}^e + \Delta y_s + \Delta y_e \tag{1}$$

$$\Delta y_s = \frac{y_T^a - y_{T-n}^a}{N} \tag{2}$$

$$\Delta y_e = k_1(y_T^a - y_{t-1}^e) \tag{3}$$

In Equation (1), $y_t^e$ denotes the extrapolated data at the current timestep $t$, while $y_{t-1}^e$ represents the extrapolated data from the previous timestep. The terms $\Delta y_s$ and $\Delta y_e$ correspond to the average variation of previously received data and the error increment between extrapolated and actual data, respectively. Equations (2) and (3) define $\Delta y_s$ and $\Delta y_e$ based on the current data point $y_T^a$ and the $n$-th previous actual data point $y_{T-n}^a$, with $N$ representing the number of extrapolated data points between $T$ and $T-n$. The parameter $k_1$ adjusts the error increment $\Delta y_e$, balancing the trade-off between curve smoothness and accuracy.

This method predicts future data points by utilizing trends from previously received data and the error between extrapolated and actual values, with minimal computational overhead. A larger $\Delta y_s$ yields smoother curves at the cost of higher error, while a smaller $\Delta y_s$ reduces error but results in rougher curves. Both $\Delta y_s$ and $\Delta y_e$ can be adjusted within a reasonable range to ensure stable and accurate extrapolation.

Figure 9 compares the raw received data with the results processed using the proposed extrapolation method and LPF. The parameters $n$ and $k_1$ are set to 1 and 0.001, respectively, while the LPF time constant is 0.01, balancing delay and curve smoothness. As shown in Fig. 9(a), while LPF achieves similar smoothness to the extrapolation method, it introduces significant delays and voltage magnitude errors, which could lead to incorrect IBR responses. In contrast, the extrapolated data provide a smoother curve with smaller errors and more accurate predictions.

Fig. 9(b) highlights the improved frequency extrapolation, where the extrapolated curve is smoother than LPF and exhibits less lag. Additionally, Fig. 9(c) demonstrates enhanced PLL-measured frequency quality, eliminating oscillations and spikes. These improvements significantly enhance the stability and reliability of distributed IBR co-simulations.

## IV. CONCLUSIONS

This paper presents the design and evaluation of real-time co-simulation communication interfaces for local and remote power system modeling. The local communication interface demonstrates high accuracy and stability in data exchange between the RTS and the MCS, with negligible communication delays for minute-level control intervals. For remote co-simulation, file-sharing and VPN-based approaches are proposed, balancing trade-offs between simplicity, latency, and security. Experimental results validate the effectiveness of these interfaces in supporting dynamic scenarios, such as load management and distributed energy resource coordination, while highlighting challenges like resolution mismatches and propagation delays.

To address instability caused by data resolution mismatches in time-sensitive co-simulations, a real-time data extrapolation method is introduced. This method significantly enhances system stability, reliability, and the overall accuracy of simulation outcomes. These findings offer valuable insights and tools for advancing real-time co-simulation in large-scale power system studies, enabling more accurate and reliable analyses of modern power grids with high integration of IBRs.

Due to space limitations, system resilience under abnormal conditions and the impact and selection of parameters for the proposed extrapolation approach are not explored in detail. Future work will focus on testing and enhancing communication interfaces for co-simulating large-scale subsystems, enhancing resilience under asynchronous events or communication disruptions, and developing strategies for compensating communication latency.

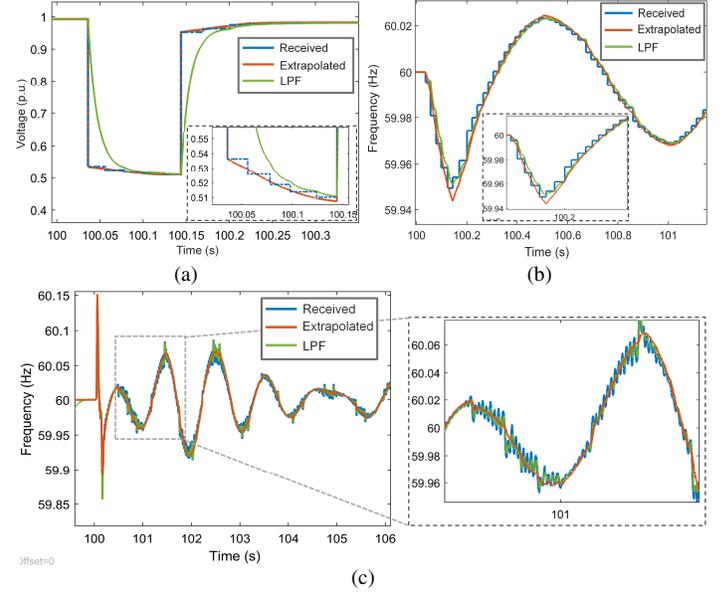

Fig. 9. Comparison of the received, extrapolated and LPF-filtered results: (a) PCC voltage, (b) frequency, (c) PLL-measured frequency at NCSU.